\documentclass[11pt]{amsart}
\usepackage{amsmath}
\usepackage{amssymb}
\usepackage{bbm}
\usepackage{a4wide}
\usepackage{sagetex}
\usepackage{tikz}

\parskip\smallskipamount

\let\set\mathbbm
\def\<#1>{\langle#1\rangle}

\def\Bold#1{\mathbf{#1}}

\newcommand\todo[1][.]{\edef\tmpa{.}\edef\tmpb{#1}%
  \ifx\tmpa\tmpb
    \typeout{To Be on page \thepage}\fbox{\bf To Be}
  \else
    \typeout{To Be on page \thepage: #1}\fbox{{\bf To Be:} #1}
  \fi
}

\begin{document}

 \author[Manuel Kauers, Maximilian Jaroschek, Fredrik Johansson]
   {Manuel Kauers\,$^\ast$, Maximilian Jaroschek\,$^\ast$, Fredrik Johansson\,$^\ast$}
 \address{Manuel Kauers, Research Institute for Symbolic Computation (RISC), J. Kepler University Linz, Austria}
 \email{mkauers@risc.uni-linz.ac.at}
 \address{Maximilian Jaroschek, Research Institute for Symbolic Computation (RISC), J. Kepler University Linz, Austria}
 \email{mjarosch@risc.uni-linz.ac.at}
 \address{Fredrik Johansson, Research Institute for Symbolic Computation (RISC), J. Kepler University Linz, Austria}
 \email{fjohanss@risc.uni-linz.ac.at}
 \thanks{$^\ast$ Supported by the Austrian FWF grant Y464-N18.}

 \title{Ore Polynomials in Sage}

 \begin{abstract}
We present a Sage implementation of Ore algebras. The main features for the most
common instances include basic arithmetic and actions; gcrd and lclm; D-finite
closure properties; natural transformations between related algebras; guessing;
desingularization; solvers for polynomials, rational functions and (generalized)
power series. This paper is a tutorial on how to use the package.
 \end{abstract}

 \maketitle


\section{Introduction}

In computer algebra, objects are often described implicitly through equations
they satisfy. For example, the exponential function $\exp(x)$ is uniquely
specified by the linear differential equation $f'(x)-f(x)=0$ and the initial
value $f(0)=1$.  Likewise, the sequence $F_n$ of Fibonacci numbers is uniquely
determined by the linear recurrence $F_{n+2}-F_{n+1}-F_n=0$ and the two initial
values $F_0=0$,~$F_1=1$.  Especially for representing functions or sequences
that cannot be expressed in ``closed form'', the differential or difference
equations they may satisfy provide an attractive way to store them on the
computer. The question is then how to calculate with objects which are given in
this form.

Algorithms for Ore algebras provide a systematic answer to this
question~\cite{bronstein96,chyzak98}.  Invented in the first half of the 20th
century~\cite{ore33} with the objective of providing a unified theory for
various kinds of linear operators, they have been used for many years in
computer algebra systems, for example in the Maple packages
OreTools~\cite{abramov03}, gfun~\cite{salvy94} or mgfun~\cite{chyzak98a}, or in
the Mathematica packages by Mallinger~\cite{mallinger96} and
Koutschan~\cite{koutschan10c,koutschan09}.

The purpose of this paper is to introduce an implementation of a collection of
algorithms related to Ore algebras for the computer algebra system
Sage~\cite{sage}. It is addressed to first-time users who are already familiar
with Sage, and with the theory of Ore algebras and its use for doing symbolic
computation related to special functions. Readers unfamiliar with Sage are referred
to~\cite{sage}, and readers unfamiliar with Ore algebras may wish to consult the
recent tutorial~\cite{kauers13} and the references given there for an introduction to
the subject.

At the time of writing, the package we describe here is still under construction
and has not yet been incorporated into the official Sage distribution. Readers
who want to try it out are invited to download the current version from 
\begin{center}
  \verb|http://www.risc.jku.at/research/combinat/software/ore_algebra|
\end{center}
and are encouraged to send us bug reports or other comments. We hope 
that the community will find the code useful.

The following instructions show how to load the code and then create an Ore
algebra~$A$ of linear differential operators and an Ore algebra~$B$ of
recurrence operators. Observe the correct application of the respective
commutation rules in both cases.

\begin{sageexample}
  sage: from ore_algebra import *
  sage: R.<x> = PolynomialRing(ZZ); A.<Dx> = OreAlgebra(R)
  sage: A
  sage: A.random_element()
  sage: Dx*x
  sage: B.<Sx> = OreAlgebra(R)
  sage: B
  sage: Sx*x
\end{sageexample}

More details on the construction of Ore algebras are given in the following
section. The construction and manipulation of elements of Ore algebras is
discussed in Section~\ref{sec:3}.

The package also supports Ore algebras with several generators already. However,
so far we offer hardly more functionality than addition and multiplication for
these. Much more functionality is available for algebras with one generator. 
Some of it is described in Section~\ref{sec:4}. We plan to add more code for the
multivariate case in the future.

\section{Ore Algebras}

An Ore algebra is determined by a base ring and a finite number of generators.
In the examples above, the base ring was $\Bold{Z}[x]$, and the generators were
$\mathrm{Dx}$ and $\mathrm{Sx}$, respectively. If no other information is provided
in the arguments, the \verb|OreAlgebra| constructor chooses the nature of the 
generators according to their name: a generator called $\mathrm{Dt}$ represents
the standard derivation $d/dt$ acting on the generator $t$ of the base ring, 
a generator called $\mathrm{Sn}$ represents the standard shift operator sending the
generator $n$ of the base ring to $n+1$. 

For this way of generating algebras, generator names must be composed of
one of the following single-letter prefixes followed by the name of a generator
of the base ring. 

\begin{center}
  \begin{tabular}{|c|c|c|}\hline
    Prefix & Name & Commutation rule \\\hline
     D & Standard derivation $d/dx$ & $\mathrm{Dx}\,x=x\,\mathrm{Dx}+1$ \\
     S & Standard shift $x\leadsto x+1$ & $\mathrm{Sx}\,x=(x+1)\,\mathrm{Sx}$ \\
     T or $\Theta$ & Eulerian derivation $x\,d/dx$ & $\mathrm{Tx}\,x=x\,\mathrm{Tx}+x$ \\
     F or $\Delta$ & Forward difference $\Delta_x$ & $\mathrm{Fx}\,x=(x+1)\mathrm{Fx}+1$ \\
     Q & $q$-shift $x\leadsto q\,x$ & $\mathrm{Qx}\,x=q\,x\,\mathrm{Qx}$ \\
     J & $q$-derivation (``Jackson derivation'') & $\mathrm{Jx}\,x=q\,x\,\mathrm{Jx}+1$ \\ 
     C & commutative generator & $\mathrm{Cx}\,x = x\,\mathrm{Cx}$ \\\hline
  \end{tabular}
\end{center}

For the $q$-shift and the $q$-derivation, the base ring must contain an element~$q$.
The element playing the role of $q$ can be specified as an optional argument. 

\begin{sageexample}
  sage: R.<x> = PolynomialRing(ZZ['q'])
  sage: A.<Qx> = OreAlgebra(R)
  sage: Qx*x
  sage: A.<Qx> = OreAlgebra(R, q=2)
  sage: Qx*x
\end{sageexample}

In general, the commutation rules of a generator~$X$ of an Ore algebra~$A$ with
base ring~$R$ are governed by two maps, $\sigma\colon R\to R$ and $\delta\colon R\to R$,
where $\sigma$ is a ring endomorphism (i.e., $\sigma(a+b)=\sigma(a)+\sigma(b)$ and
$\sigma(ab)=\sigma(a)\sigma(b)$ for all $a,b\in R$) and $\delta$ is a skew-derivation
for $\sigma$ (i.e., $\delta(a+b)=\delta(a)+\delta(b)$ and $\delta(ab)=\delta(a)b
+\sigma(a)\delta(b)$ for all $a,b\in R$). With two such maps being given, the
generator $X$ satisfies the commutation rule $Xa=\sigma(a)X+\delta(a)$ for every
$a\in R$. If there is more than one generator, then each of them has its own pair
of maps $\sigma,\delta$. Different generators commute with each other; 
noncommutativity only takes place between generators and base ring elements. 

It is possible to create an Ore algebra with user specified commutation rules.
In this form, each generator must be declared by a tuple $(X,\sigma,\delta)$, 
where $X$ is the name of the generator (a string), and $\sigma$ and $\delta$
are dictionaries which contain the images of the generators of the base
ring under the respective map. Here is how to specify an algebra of difference
operators in this way:

\begin{sageexample}
  sage: R.<x> = ZZ['x']
  sage: A = OreAlgebra(R, ('X', {x:x+1}, {x:1}))
  sage: X = A.gen()
  sage: X*x
\end{sageexample}

As another example, here is how to define an algebra of differential operators 
whose base ring is a differential field $K=\set Q(x,y,z)$ where $y$ represents
$\exp(x)$ and $z$ represents~$\log(x)$:

\begin{sageexample}
  sage: K = ZZ['x','y','z'].fraction_field()
  sage: x,y,z = K.gens()
  sage: A = OreAlgebra(K, ('D', {}, {x:1, y:y, z:1/x}))
  sage: D = A.gen()
  sage: D*x, D*y, D*z
\end{sageexample}

In the dictionary specifying~$\sigma$, omitted generators are understood to be
mapped to themselves, so that \verb|{}| in the definition of $A$ in the example
above is equivalent to \verb|{x:x,y:y,z:z}|. In the dictionaries
specifying~$\delta$, omitted generators are understood to be mapped to zero.

For Ore algebras with several generators, it is possible to mix specifications
of generators via triples $(X,\sigma,\delta)$ with generators using the naming
convention shortcuts as explained before. Continuing the previous example,
here is a way to define an algebra $A$ over $K$ with two generators, a $D$ 
that behaves like before, and in addition an $Sx$ which acts like the standard
shift on~$x$ and leaves the other generators fixed.

\begin{sageexample}
  sage: A = OreAlgebra(K, ('D', {}, {x:1, y:y, z:1/x}), 'Sx')
  sage: D, Sx = A.gens()
  sage: D*x, Sx*x
  sage: D*y, Sx*y
  sage: D*z, Sx*z
\end{sageexample}

In theory, any integral domain can serve as base ring of an Ore algebra. Not so
in our implementation. Here, base rings must themselves be polynomial rings
(univariate or multivariate), or fraction fields of polynomial rings. Their base
rings in turn may be either $\set Z$, $\set Q$, a prime field $GF(p)$, or a
number field $\set Q(\alpha)$, or --- recursively --- some ring which itself
would be suitable as base ring of an Ore algebra.

\begin{sageexample}
  sage: ZZ['x'].fraction_field()['y','z'] ### OK
  sage: GF(1091)['x','y','z']['u'] ### OK
  sage: ( ZZ['x','y','z'].quotient_ring(x^2+y^2+z^2-1) )['u'] ### not OK
  sage: GF(9, 'a')['x'] ### not OK
\end{sageexample}

Note that the maps $\sigma$ and $\delta$ must leave all the elements of
the base ring's base ring fixed. They may only have nontrivial images for
the top level generators. 

The constituents of an Ore algebra $A$ can be accessed through the methods
summarized in the following table. Further methods can be found in the 
documentation. 

\begin{center}
  \begin{tabular}{|l|p{.5\hsize}|}
    \hline
    Method name & short description \\\hline
    \verb|associated_commutative_algebra()| & returns a polynomial ring with the
       same base ring as $A$ and whose generators are named like the generators
       of~$A$\\
    \verb|base_ring()| & returns the base ring of $A$\\
    \verb|delta(i)| & returns a callable object representing the delta map
       associated to the $i$th generator (default: $i=0$) \\
    \verb|gen(i)| & returns the $i$th generator (default: $i=0$)\\
    \verb|sigma(i)| & returns a callable object representing the sigma map
       associated to the $i$th generator (default: $i=0$) \\
    \verb|var(i)| & returns the name of the $i$th generator (default: $i=0$)\\\hline
  \end{tabular}
\end{center}

\smallskip\goodbreak

Examples: 

\begin{sageexample}
  sage: R.<x> = ZZ['x']; A.<Dx> = OreAlgebra(R)
  sage: A
  sage: A.associated_commutative_algebra()
  sage: A.base_ring()
  sage: A.gen()
  sage: s = A.sigma(); d = A.delta(); 
  sage: s(x^5), d(x^5)
\end{sageexample}

\section{Ore Polynomials}\label{sec:3}

Ore polynomials are elements of Ore algebras, i.e., Sage objects whose parent is an Ore algebra
object as described in the previous section. They can be constructed by addition and multiplication
from generators and elements of the base ring. 

\begin{sageexample}
  sage: R.<x> = ZZ['x']; A.<Dx> = OreAlgebra(R)
  sage: (5*x^2+3*x-7)*Dx^2 + (3*x^2+8*x-1)*Dx + (9*x^2-3*x+8)
\end{sageexample}

Alternatively, an Ore polynomial can be constructed from any piece of data that is also accepted
by the constructor of the associated commutative algebra. The associated commutative algebra of
an Ore algebra is the commutative polynomial ring with the same base ring as the Ore algebra 
and with generators that are named like the generators of the Ore algebra. In particular, it is
possible to create an Ore polynomial from the corresponding commutative polynomial, from a 
coefficient list, or even from a string representation.

\begin{sageexample}
  sage: R.<x> = ZZ['x']; A.<Dx> = OreAlgebra(R)
  sage: Ac = A.associated_commutative_algebra()
  sage: Ac
  sage: A(Ac.random_element())
  sage: A([5*x,7*x-3,3*x+1])
  sage: A("(5*x^2+3*x-7)*Dx^2 + (3*x^2+8*x-1)*Dx + (9*x^2-3*x+8)")
\end{sageexample}

Ore polynomials can also be created from Ore polynomials that belong to other algebras, provided
that such a conversion is meaningful. 

\begin{sageexample}
  sage: R.<x> = ZZ['x']; A.<Dx> = OreAlgebra(R)
  sage: L = (5*x^2+3*x-7)*Dx^2 + (3*x^2+8*x-1)*Dx + (9*x^2-3*x+8)
  sage: L.parent()
  sage: B = OreAlgebra(QQ['x'], 'Dx')
  sage: L = B(L)
  sage: L.parent()
\end{sageexample}

In accordance with the Sage coercion model, such conversions take place automatically (if possible) 
when operators from different algebras are added or multiplied. Note that the result of such an operation
need not belong to either of the parents of the operands but may instead have a suitable ``common extension'' 
as parent. 

\begin{sageexample}
  sage: A = OreAlgebra(ZZ['t']['x'], 'Dx')
  sage: B = OreAlgebra(QQ['x'].fraction_field(), 'Dx')
  sage: L = A.random_element() + B.random_element()
  sage: L.parent()
\end{sageexample}

\section{Selected Methods}\label{sec:4}

Besides basic arithmetic for Ore operators, the package provides a wide range of
methods to create, manipulate and solve several different kinds of operators.
Some of these methods are accessible in any Ore algebra while others are tied
specifically to, e.g., recurrence operators or differential operators. 

In this section, we give an overview of the functionality provided by
the package. Because of space limitation, only some of the available methods
can be discussed here. For further information, we refer to the documentation. 

\subsection{Methods for General Algebras}

A univariate Ore algebra over a field is a left Euclidean domain, which means
that it is possible to perform left division with remainder. Building upon this,
the greatest common right divisor (GCRD) and the least common left multiple
(LCLM) of two Ore polynomials can be computed. The package provides a number of
methods to carry out these tasks.

\begin{center}
  \begin{tabular}{|l|p{.65\hsize}|}
    \hline
    Method name & short description \\\hline
    \verb|A.quo_rem(B)| & returns the left quotient and the left remainder of $A$ and $B$.\\
    \verb|A.gcrd(B)| & returns the greatest common right divisor of $A$ and $B$.\\
    \verb|A.xgcrd(B)| & returns the greatest common right divisor of $A$ and $B$ and the according B\'ezout coefficients.\\
    \verb|A.lclm(B)| & returns the least common left multiple of $A$ and $B$.\\
    \verb|A.xlclm(B)| & returns the least common left multiple $L$ of $A$ and $B$ and the left quotients of $L$ and $A$ and of $L$ and $B$.\\
    \verb|A.resultant(B)| & returns the resultant of $A$ and $B$ (see \cite{li96} for its definition and
    properties).\\
    \hline
  \end{tabular}
\end{center}

All these methods are also available for Ore operators living in univariate Ore
algebras over a base ring $R$ which does not necessarily have to be a
field. The operators are then implicitly assumed to live in the respective Ore
algebra over the quotient field $K$ of~$R$. The output will be the
GCRD (LCLM, quotient, remainder) in the Ore algebra over $K$ but not over
$R$, in which these objects might not exist or might not be computable.

\begin{sageexample}
  sage: A = OreAlgebra(ZZ['n'], 'Sn')
  sage: G = A.random_element(2)
  sage: L1, L2 = A.random_element(7), A.random_element(5)
  sage: while L1.gcrd(L2) != 1: L2 = A.random_element(5)                       
  sage: L1, L2 = L1*G, L2*G                                                        
  sage: L1.gcrd(L2) == G.normalize()
  sage: L3, S, T = L1.xgcrd(L2)                             
  sage: S*L1 + T*L2 == L3
  sage: LCLM = L1.lclm(L2)
  sage: LCLM 
  sage: LCLM.order() == L1.order() + L2.order() - G.order()
\end{sageexample}

The GCRD is only unique up to multiplication (from the left) by an element from
the base ring. The method \verb|normalize| called in line~6 of the listing above
multiplies a given operator from the left by some element from the base ring
such as to produce a canonical representative from the class of all the operators
that can be obtained from each other by left multiplication of a base ring
element. This facilitates the comparison of output.

The efficiency of computing the GCRD depends on the size of the
coefficients of intermediate results, and there are different strategies to
control this growth via so called polynomial remainder sequences (PRS). The default is
the improved PRS described in~\cite{jaroschek13b}, which will usually be the
fastest choice. Other strategies can be selected by the option \verb|prs|.

\begin{sageexample}
  sage: A = OreAlgebra(ZZ['n'], 'Sn')
  sage: L1, L2 = A.random_element(3), A.random_element(2)
  sage: algs = ["improved", "classic", "monic", "subresultant"]
  sage: [L1.gcrd(L2, prs=a) for a in algs]
\end{sageexample}

If $L_1,L_2$ are operators, then the solutions of their GCRD are precisely the
common solutions of $L_1$ and~$L_2$. The LCLM, on the other hand, is the minimal
order operator whose solution space contains all the solutions of $L_1$ and
all the solutions of~$L_1$. Because of this property, GCRD and LCLM are
useful tools for constructing operators with prescribed solutions. For example,
here is how to construct a differential operator which has the solutions $x^5$
and $\exp(x)$, starting from the obvious operators annihilating $x^5$
and~$\exp(x)$, respectively.

\begin{sageexample}
  sage: R.<x> = ZZ[]; A.<Dx> = OreAlgebra(R)
  sage: L = (Dx - 1).lclm(x*Dx - 5)
  sage: L
  sage: L(x^5)
  sage: L(exp(x)).full_simplify() 
\end{sageexample}

Observe how in the last two lines we apply the operator~$L$ to other
objects. Such applications are not defined for every algebra and in general have
to be specified by the user through an optional argument:

\begin{sageexample}
  sage: A.<Qqn> = OreAlgebra(ZZ['q']['qn'])
  sage: var('q', 'n', 'x')
  sage: (Qqn^2+Qqn+1)(q^n, action=lambda expr: expr.substitute(n=n+1))
  sage: (Qqn^2+Qqn+1)(x, action=lambda expr: expr.substitute(x=q*x))
\end{sageexample}

Thanks to the LCLM operation discussed above, we have the property that when $f$
and $g$ are two objects which are annihilated by some operators $L_1,L_2$
belonging to some Ore algebra~$A$ then this algebra contains also an operator
which annihilates their sum~$f+g$. In other words, the class of solutions of operators of
$A$ is closed under addition. It turns out that similar closure properties hold
for other operations. The following table lists some of the corresponding
methods. Methods for more special closure properties will appear further below.

\begin{center}
\begin{tabular}{|l|p{.55\hsize}|}
\hline
Method name & short description \\\hline
\verb|lclm()| & computes an annihilating operator for $f+g$ from annihilating operators for $f$ and $g$ \\
\verb|symmetric_product()| & computes an annihilating operator for $fg$ from annihilating operators for $f$ and $g$ \\
\verb|symmetric_power()| & computes an annihilating operator for $f^n$ from an annihilating operator for $f$ and a given positive integer~$n$\\ 
\verb|annihilator_of_associate()| & computes an annihilating operator for $M(f)$ from an annihilating operator for $f$ and a given operator~$M$\\ 
\verb|annihilator_of_polynomial()| & computes an annihilating operator for the object $p(f,\partial f,\partial^2 f,\dots)$ from an annihilating operator for $f$ and a given multivariate polynomial~$p$. \\
\hline
\end{tabular}
\end{center}

As an example application, let us prove Cassini's identity for Fibonacci
numbers: 
\[
  F_{n+1}^2 - F_nF_{n+2}=(-1)^n.
\]
The idea is to derive, using commands from the table above, a recurrence
satisfied by the left hand side, and then show that this recurrence is also
valid for the right hand side.

\begin{sageexample}
  sage: A.<Sn> = OreAlgebra(ZZ['n'])
  sage: fib = Sn^2 - Sn - 1
  sage: R.<x0,x1,x2> = ZZ['n']['x0','x1','x2']
  sage: fib.annihilator_of_polynomial(x1^2 - x0*x2) 
\end{sageexample}

As this operator obviously annihilates $(-1)^n$, the proof is complete after
checking that the identity holds for $n=0$. Another way of carrying out the
same computation using the other commands would be as follows.

\begin{sageexample}
  sage: A.<Sn> = OreAlgebra(ZZ['n'])
  sage: fib = Sn^2 - Sn - 1
  sage: L1 = fib.annihilator_of_associate(Sn).symmetric_power(2)
  sage: L2 = fib.annihilator_of_associate(Sn^2).symmetric_product(fib)
  sage: L1.lclm(L2)
\end{sageexample}

Observe that the resulting operator again annihilates~$(-1)^n$, but its order is
higher than the operator obtained before, so we need to check more initial
values to complete the proof. For larger computations, the command
\verb|annihilator_of_polynomial| would also consume less computation time than
the step-by-step approach.

\subsection{Methods for Special Algebras}

For the elements of some of the most important algebras, additional methods have
been implemented. The following table lists some of the additional methods
available for differential operators, i.e., elements of an Ore algebra of the
form~$R[x]\<Dx>$ or $K(x)\<Dx>$.

\begin{center}
  \begin{tabular}{|l|p{.55\hsize}|}
    \hline
    Method name & short description \\\hline
    \verb|to_S()| & converts to a recurrence operator for the Taylor series solutions at the origin \\
    \verb|to_F()| & converts to a difference operator for the Taylor series solutions at the origin \\
    \verb|to_T()| & rewrites in terms of the Euler derivative \\
    \verb|annihilator_of_integral()| & converts an annihilator for $f(x)$ to one for $\int f(x) dx$\\
    \verb|annihilator_of_composition()| & converts an annihilator for $f(x)$ to one for $f(a(x))$ where $a(x)$ is algebraic over the base ring\\
    \verb|desingularize()| & computes a left multiple of this operator with polynomial coefficients and 
           lowest possible leading coefficient degree\\
    \verb|associate_solutions(p)| & applied to an operator~$P$, this computes, if possible, an operator $M$
    and a rational function $m$ such that $DM=p+mP$ (see \cite{abramov99} for further information)\\
    \verb|polynomial_solutions()| & computes the polynomial solutions of this operator\\
    \verb|rational_solutions()| & computes the rational function solutions of this operator\\
    \verb|power_series_solutions()| & computes power series solutions of this operator\\
    \verb|generalized_series_solutions()| & computes generalized series solutions of this operator \\ \hline
  \end{tabular}
\end{center}

\smallskip

As an example application, we compute an annihilator for the error function
$\frac{2}{\sqrt{\pi}} \int_0^x \exp(-t^2) dt$, starting from the differential
equation for $\exp(x)$, and produce the recurrence for the Taylor series
coefficients at the origin. Finally, we compute the series solutions of the
differential operator at infinity. 

\begin{sageexample}
  sage: R.<x> = ZZ['x']; A.<Dx> = OreAlgebra(R, 'Dx')                            
  sage: (Dx - 1).annihilator_of_composition(-x^2)
  sage: L = (Dx + 2*x).annihilator_of_integral()                                     
  sage: L
  sage: L.to_S(OreAlgebra(ZZ['n'], 'Sn'))                          
  sage: L.power_series_solutions(10)                               
  [x - 1/3*x^3 + 1/10*x^5 - 1/42*x^7 + 1/216*x^9 + O(x^10), 1 + O(x^10)]
  sage: L.annihilator_of_composition(1/x).generalized_series_solutions()
\end{sageexample}

The last output implies that the operator annihilating $\int_0^x \exp(-t^2) dt$
also admits a solution which behaves for $x\to\infty$ like $\frac1x\exp(-x^2)$.

The next example illustrates the methods for finding rational and polynomial
solutions of an operator~$L$.  These methods accept as an optional parameter an
inhomogeneous part consisting of a list (or tuple) of base ring elements,
$(f_1,\dots,f_r)$. They return as output a list of tuples $(g,c_1,\dots,c_r)$
with $L(g)=c_1f_1+\cdots+c_rf_r$ where $g$ is a polynomial or rational function
and $c_1,\dots,c_r$ are constants, i.e., elements of the base ring's base
ring. The tuples form a vector space basis of the solution space.

In the example session below, we start from two polynomials $p,q$, then compute an 
operator~$L$ having $p$ and $q$ as solutions, and then recover $p$ and $q$ from~$L$.
Note that for consistency also the solutions of homogeneous equations are returned as 
tuples. At the end we give an example for solving an inhomogeneous equation.

\begin{sageexample}
  sage: R.<x> = ZZ[]
  sage: p = x^2 + 3*x + 8; q = x^3 - 7*x + 5
  sage: A.<Dx> = OreAlgebra(R)
  sage: L = (p*Dx - p.derivative()).lclm(q*Dx - q.derivative())
  sage: L
  sage: L.polynomial_solutions()
  sage: M = (2*x+3)*Dx^2 + (4*x+5)*Dx + (6*x+7)
  sage: sol = M.polynomial_solutions([1,x,x^2,x^3])
  sage: sol
  sage: map(lambda s: M(s[0]) == s[1]+s[2]*x+s[3]*x^2+s[4]*x^3, sol)
\end{sageexample}

The functions \verb|polynomial_solutions| and \verb|rational_solutions| are not
only defined for differential operators but also for recurrence operators, i.e.,
elements of an Ore algebra of the form $R[x]\<Sx>$ or~$K(x)\<Sx>$. Some other
methods defined for recurrence operators are listed in the following table. 

\begin{center}
  \begin{tabular}{|l|p{.55\hsize}|}
    \hline
    Method name & short description \\\hline
    \verb|to_D()| & converts annihilator for the coefficients in a power series to a
        differential operator for the sum \\
    \verb|to_F()| & converts shift operator to a difference operator\\
    \verb|to_T()| & converts to a differential operator in terms of the Euler derivative \\
    \verb|annihilator_of_sum()| & converts an annihilator for $f(n)$ to one for the sum $\sum_{k=0}^n f(k)$\\
    \verb|annihilator_of_composition()| & converts an annihilator for $f(n)$ to one for $f(\lfloor un+v \rfloor)$ where $u, v \in \set Q$ \\
    \verb|annihilator_of_interlacing()| & interlaces two or more sequences\\
    \verb|desingularize()| & computes a left multiple of this operator with polynomial coefficients and 
           lowest possible leading coefficient degree\\
    \verb|associate_solutions(p)| & applied to an operator~$P$, this computes, if possible, an operator $M$
    and a rational function $m$ such that $(S-1)M=p+mP$ (see \cite{abramov99} for further information)\\
    \verb|polynomial_solutions()| & computes the polynomial solutions of this operator\\
    \verb|rational_solutions()| & computes the rational function solutions of this operator\\
    \verb|generalized_series_solutions()| & computes asymptotic expansions of sequences annihilated by the operator\\ 
    \verb|to_list()| & computes terms of a sequence annihilated by the operator\\\hline
  \end{tabular}
\end{center}

\smallskip

As an example application, we compute an annihilator for the sequence
$c(n) = \sum_{k=0}^n 1 / k!$:

\begin{sageexample}
  sage: R.<n> = ZZ[]; A.<Sn> = OreAlgebra(R)
  sage: inverse_factorials = (n + 1) * Sn - 1
  sage: partial_sums = inverse_factorials.annihilator_of_sum()
  sage: partial_sums
\end{sageexample}

The \verb|to_list| method returns the first few values of a sequence,
given the initial values:

\begin{sageexample}
  sage: L = partial_sums.to_list([1, 2], 8)
  sage: L
  sage: N(L[7])
\end{sageexample}

We compute the asymptotic expansion of the sequence of terms
to estimate how many terms we need to approximate $e$ to a given
number of digits:

\begin{sageexample}
  sage: digits = 10^5
  sage: asymp = inverse_factorials.generalized_series_solutions(3)
  sage: target = lambda x: log(abs(asymp[0](RR(x))), 10) + digits
  sage: num_terms = ceil(find_root(target, 1, 10^6))
  sage: num_terms
\end{sageexample}

In some cases, for example when the base ring is $\mathbb{Z}$ or
$\mathbb{Z}[x]$, isolated values of a sequence can be
computed asymptotically faster for large $n$ than by listing all
values, using the binary splitting technique.
The \verb|forward_matrix_bsplit| method, called with argument~$n$,
returns a matrix $P$ and a polynomial $Q$ such that $P / Q$ multiplied by a
column vector of initial values $c_0, c_1, \ldots$
yields $c_n, c_{n+1}, \ldots$.
This way, computing $10^5$ digits of $e$ takes a fraction of a second:

\begin{sageexample}
  sage: e_approx = N(e, 400000)
  sage: P, Q = partial_sums.forward_matrix_bsplit(num_terms)
  sage: u = Matrix([[e_approx], [e_approx]]) - P * Matrix([[1], [2]]) / Q
  sage: u.change_ring(RealField(20))
\end{sageexample}

\subsection{Guessing}

Guessing is, in some sense, the reverse operation of \verb|to_list| for
recurrence operators, or of \verb|power_series_solutions| for differential
operators.  It is of the most popular features of packages like gfun, and there
are even some special purpose packages dedicated to this
technique~\cite{kauers09a,hebisch11}. The basic idea is simple. Given
a finite array of numbers, thought of as the first terms of an infinite
sequence, we want to know whether this sequence satisfies a recurrence. The
algorithm behind a guessing engine searches for small equations matching the
given data.  Generically, no such equations exist, so if some are found, it is
fair to ``guess'' that they are in fact valid equations for the whole infinite
sequence.

We provide a guessing function which takes as input a list of terms and an Ore
algebra, and returns as output an operator which matches the given data and
which, in some measure, would be unlikely to exist for random data.

\begin{sageexample}
  sage: data = [ 0, 1, 1, 2, 3, 5, 8, 13, 21, 34, 55 ]
  sage: L = guess(data, OreAlgebra(ZZ['n'], 'Sn'))
  sage: L
  sage: L(data)
  sage: M = guess(data, OreAlgebra(ZZ['x'], 'Dx'))
  sage: M
  sage: M(x/(1-x-x^2))
\end{sageexample}

If an algebra of differential operators is supplied as second argument, the data
is understood as the first few coefficients of a power series. The output
operator is expected to have this power series as solution.

It can happen that the procedure is unable to find an operator matching the 
given data. In this case, an exception is raised. There are two possible 
explanations for such an event. Either the sequence in question does not 
satisfy any equations, or it does but the equations are so big that more
data is needed to detect them. 

Several options are available for customizing the search for relations.  In
order to explain them, we first need to give some details on the underlying
algorithms. For simplicity of language, we restrict here to the case of
recurrence operators. The situation for differential operators is very similar.

For the most typical situations, there are two important hyperbolas. One
describes the region in the $(r,d)$-plane consisting of all points for which
there exists an operator of order~$r$ and degree~$d$ truly satisfied by the
sequence in question. (See~\cite{jaroschek13a} for an explanation why the
boundary of this region is usually a hyperbola.) The second describes the region
of all points $(r,d)$ for which an operator of order~$r$ and degree~$d$ can be
detected when $N$ terms are provided as input. This region is determined by the
requirement $(r+1)(d+2)<N$.

The method tests a sequence of points $(r_1,d_1)$, $(r_2,d_2)$, \dots\ right
below this second hyperbola. Success at a point $(r_i,d_i)$ means that some
evidence for an operator of order $\leq r_i$ and degree $\leq d_i$ has been found. 
This operator however is not explicitly computed. Instead, the method uses the
partial information found about this operator to calculate an operator which with 
high probability is the minimal order operator satisfied by the sequence in question. 
This operator is usually more interesting than the one at $(r_i,d_i)$, and its 
computation is usually more efficient. 

Using the option \verb|path|, the user can specify a list of points $(r_i,d_i)$
which should be used instead of the standard path. By setting the options \verb|min_degree|,
\verb|max_degree|, \verb|min_order|, \verb|max_order|, all points $(r,d)$ of the path
are discarded for which $r$ or $d$ is not within the specified bounds. These options
can be used to accelerate the search in situations where the user has some knowledge
(or intuition) about the size of the expected equations. 

\hangindent=-6.3cm\hangafter=-13\leavevmode
\smash{\raise-5.3cm\rlap{\kern\hsize\kern-5.1cm\begin{tikzpicture}[scale=.45]
  \draw[->] (10.5,0) node {$r$} (-.2,0) -- (10, 0) ;
  \draw[->] (0,10.5) node {$d$} (0,-.2) -- (0, 10) ;
  \draw (1.2, 3.5) -- (1.4, 4) (1.2, 3) -- (1.6, 4) (1.2, 2.5) -- (1.8, 4) (1.2, 2) -- (2, 4)
        (1.4, 2) -- (2.2, 4) (1.6, 2) -- (2.4, 4) (1.8, 2) -- (2.6, 4) (2, 2) -- (2.8, 4) 
        (2.2, 2) -- (3, 4) (2.4, 2) -- (3.2, 4) (2.6, 2) -- (3.4, 4) (2.8,2)--(3.6,4) (3,2)--(3.8,4)
        (3.2, 2) -- (4, 4) (3.4, 2) -- (4.2, 4) (3.6, 2) -- (4.4, 4) (3.8,2)--(4.6,4) (4,2)--(4.8,4)
        (4.2, 2) -- (5, 4) (4.4, 2) -- (5.2, 4) (4.6, 2) -- (5.4, 4) (4.8,2)--(5.6,4) (5,2)--(5.8,4); 
  \fill[lightgray] (.75, 10) .. controls (1.5,2.5) and (2.5,1.5) .. (10, .75) -- (10, 10) -- (.75, 10);
  \draw[thick] (2, 10) .. controls (2, 1.2) and (2, 1.2) .. (10, 1.2); 
  \draw (-.2, 2) node {\vbox{\llap{\footnotesize min }\kern-3pt\llap{\footnotesize degree }}} -- (10, 2) 
        (-.2, 4) node {\vbox{\llap{\footnotesize max }\kern-3pt\llap{\footnotesize degree }}} -- (10, 4);
  \draw (1.2, -.2) node {\hbox to0pt{\rule{0pt}{2em}\hss\footnotesize min order\hss}} -- (1.2, 10)
        (5, -.2) node {\hbox to0pt{\rule{0pt}{2em}\hss\footnotesize max order\hss}} -- (5, 10);
  \fill (2.5, 3) circle(4pt) node[right] {\rule[-1em]{0pt}{0pt}$A$} (2, 9) circle(4pt) node[right] {$B$} ;
\end{tikzpicture}}}%
The figure on the right illustrates the typical situation for guessing problems
that are not too small and not too artificial. The gray region indicates
the area which is not accessible with the given amount of data. Only the points
$(r,d)$ below it can be tested for an operator of order~$r$ and degree~$d$ that
fits to the given data. Let's assume that operators exist on and above the solid
black hyperbola. The user will usually not know this curve in advance but may
have some expectations about it and can restrict the search accordingly, for
example to the dashed area shown in the figure. The method will detect the
existence of an operator, say at point~$A$, and construct from the information
gained at this point an operator of minimal possible order, which may correspond
to point~$B$.  This operator is returned as output. Note that the degree of the
output may exceed the value of \verb|max_degree|, and its order may be smaller
than $\verb|min_order|$:

\begin{sageexample}
sage: data = [(n+1)^10*2^n + 3^n for n in xrange(200)]
sage: L = guess(data, OreAlgebra(ZZ['n'],'Sn'), min_order=3, max_degree=5)
sage: L.order(), L.degree()
\end{sageexample}

In order to test a specific point $(r,d)$, the data array must contain at least $(r+1)(d+2)$ terms. 
If it has more terms, the guess becomes more trustworthy, but also the computation time increases.
By setting the option \verb|ensure| to a positive integer~$e$, the user can request that only 
such points $(r,d)$ should be tested for which the data array contains at least $e$ more terms than
needed. This increases the reliability. By setting the option \verb|cut| to a positive integer~$c$,
the user requests that for testing a point $(r,d)$, the method should take into account at most $c$
more terms than needed. If the data array contains more terms, superfluous ones are ignored in the
interest of better performance. We must always have $0\leq e\leq c\leq \infty$. The default setting
is $e=0$, $c=\infty$. 


\end{document}